\documentclass[a4paper]{article}
\usepackage{graphicx}
\usepackage{twocolpceurws}
\usepackage{amsmath}
\usepackage{amsthm}
\usepackage{listings}
\usepackage{multirow}
\usepackage{multicol}
\usepackage[table]{xcolor}
\usepackage{url}

\usepackage{todonotes}

\usepackage{authblk}

\newtheorem{definition}{Definition}
\newtheorem{example}{Example}
\newcommand\pms{\textit{pms}}
\definecolor{truepositive}{HTML}{FE0000}
\definecolor{falsepositive}{HTML}{FFCCC9}

\title{Towards Automated Android App Collusion Detection}

\author[1]{Irina Mariuca As\u avoae} 
\author[2]{Jorge Blasco}
\author[2]{Thomas M. Chen}
\author[3]{Harsha Kumara Kalutarage}
\author[4]{Igor Muttik}
\author[5]{Hoang Nga Nguyen}
\author[$\ $,1]{Markus Roggenbach \thanks{$\ \ $Corresponding author \texttt{M.Roggenbach@swan.ac.uk}}}
\author[5]{Siraj Ahmed Shaikh}

\affil[1]{Swansea University, UK}
\affil[2]{City University London, UK}
\affil[3]{Queen's University of Belfast, UK}
\affil[4]{Intel Security, UK}
\affil[5]{Coventry University, UK}

\institution{}

\begin{document}

\maketitle

\begin{abstract}
Android OS supports multiple communication methods between apps.
%
%% The Android OS allows to transmit sensitive data from one app (in a
%% sandbox with permissions to handle such data) into another app (in
%% another sandbox which has been denied permission to handle such
%% data).
%
This opens the possibility to carry out threats in a collaborative
fashion, c.f.\ the Soundcomber example from 2011. In this paper we
provide a concise definition of collusion and report on a number of
automated detection approaches, developed in co-operation with Intel
Security.
\end{abstract}

\section{Introduction}

The Android operating system (OS)
%\textregistered
%\textsuperscript{TM}
is designed with a number of
built-in security features such as app sandboxing and fairly granular
access controls based on permissions. In real life, however, isolation
of apps is limited. In some respects even the opposite is true --
there are many ways for apps to communicate across sandbox
boundaries. The Android OS supports multiple communication methods
which are fully documented (such as messaging via intents). The ability of apps with different security postures to communicate has a negative effect on security as an app (in a sandbox which has permissions to handle such data) is allowed to let sensitive data flow to another app (in another sandbox which has been denied permission to handle such data) and eventually leak out.
 
The Android ecosystem exacerbates the problem as the market pressure leads many developers to embed advertisement libraries into their
apps. As a result, such code may be present in thousands of apps (all bearing different permissions). Advertisers have a known tendency to disregard user privacy in favour of monetisation. So there exists a risk that ad-libraries may
communicate between sandboxes – even without knowledge of apps’
authors – to transmit sensitive data across, risking exposure and disregarding privacy. 

Of course, any unscrupulous developer may also split functionality which they prefer to hide between multiple apps. Malicious behaviours similar to this are evident from known cases of apps exploiting insecure exposure of sensitive data by other apps \cite{arzt2014flowdroid}.
 
Researchers have demonstrated that sets of apps may violate the permissions model causing data leaks or carrying malware
\cite{schlegel2011soundcomber}. Such apps are called {\sl colluding
  sets of apps} and the phenomenon is called {\sl app
  collusion}. Unfortunately, there are no effective tools to detect for app collusion. The search space posed by possible combination of apps means that this is not straightforward. Effective methods are needed to narrow down the search to collusion candidates of interest. 
  
  This paper contributes towards a practical automated system for collusion detection. We give a definition of collusion in Section
\ref{sec:def}. This is followed by two potential approaches to filter down to potential candidates for collusion, using a rule based approach developed in Prolog in Section \ref{sec:prolog} and another statistical approach in Section \ref{sec:coventry}. Section \ref{sec:swansea} presents a model-checking approach to detecting collusion in Android Dalvik apps. Section \ref{sec:results} delves into the experimental outcomes and Section \ref{sec:related} discusses related work. Section \ref{sec:summary} summarises our contributions and Section \ref{sec:futurework} concludes the paper with thoughts on future work.

\section{Defining collusion}\label{sec:def}

Our notion of collusion refers to the ability for a set of
apps to carry out a threat in a collaborative fashion. This is in
contrast to most existing work, where collusion is usually associated
with inter-app communications and information leakage (see Section
\ref{sec:related}). We consider that colluding apps can
carry out any threat such as the ones posed by single apps. The range of such threats includes~\cite{suarez2014evolution}:
\begin{itemize}
\item Information theft: happens when one app accesses sensitive
  information and the other sends information outside the device
  boundaries.
\item Money theft: happens when an app sends information to
  another app that is capable of using money sensitive API
  calls (e.g. SMS).
\item Service misuse: happens when one app is able to control
  a system service and receives information (commands) from another
  app to control those services.
\end{itemize}

A threat can be described by a set of actions executed in a certain
order. We model this by a partially ordered set $(T,\leq),$ where $T$
is the set of actions and $\leq$ specifies the execution order. When
$(T,\leq)$ is carried out, actions from $T$ are sequentially executed,
according to some total order $\leq^*$ such that $\leq \subseteq
\leq^*$; in other words, $(T,\leq^*)$ is a total extension of
$(T,\leq)$. Let $Ex((T,\leq))$ denote the set of all possible total
extensions of $(T,\leq)$; i.e., all possible ways of carrying out 
threat $(T,\leq)$. %% We have $Ex((T,\leq)) = \{ (T,\leq^*)
%% \mid\,\leq \subseteq \leq^* \land \leq^* \text{ is total} \}$. To this
%% end, a sequence of actions can be seen interchangeably as a totally
%% order set. 
%% Furthermore, one may obfuscate a total extension of a
%% threat by scattering it with meaningless actions. However, the total
%% extension must be a subsequence\footnote{A sequence $a_1\ldots a_n$ is
%%   a subsequence of another one $b_1\ldots b_m$ iff $a_1\ldots a_n$ can
%%   be obtained from $b_1\ldots b_m$ by deleting some elements $b_i$'s.}
%% of the execution.
Similarly, we also define inter-app communication as a partially
ordered set. 

We define collusion as follows:
\begin{itemize} 

\item[A1:] Actions are operations provided by Android API (such as record
audio, access file, write file, send data, etc.). Let $Act$ denote the set of
all actions.

\item[A2:] Actions can be characterised by a number of 
%
%(static/  dynamic) 
%
attributes (such as permissions, input parameters, etc.). Let $B$
denote the set of all action attributes and $\pms:Act \to \wp(B)$
specify the set of permissions required by Android to execute an
action.

\item[A3:] A threat $t = (T,\leq)$ is a partially ordered set. Let
  $\tau$ denote the set of all threats. In the scope of this paper,
  $\tau$ represents the set of all known threats caused by single
  applications.

\item[A4:] An inter-app communication $c = (C,\leq)$ is a partially
  ordered set. Let $com$ denote the set of all known inter-app
  communications.

\end{itemize}

\begin{definition}[Collusion]\label{def:collusion}
A set $S$ consisting of at least two apps is \emph{colluding}
if together they execute a sequence $A \in Act^*$ such that:
\begin{enumerate}

\item there exists a subsequence $A'$ of $A$ such that $A' \in Ex(t)$
  for some $t \in \tau$; furthermore, $A'$ is collectively executed
  involving all apps in $S$, i.e., each app in $S$
  executes at least one action in $A'$; and

\item there exists a subsequence $C'$ of $A$ such that $C' \in Ex(c)$
  for some $c \in com$.

\end{enumerate} 

\end{definition}

\begin{figure}
\centering
\includegraphics[height=1.6cm]{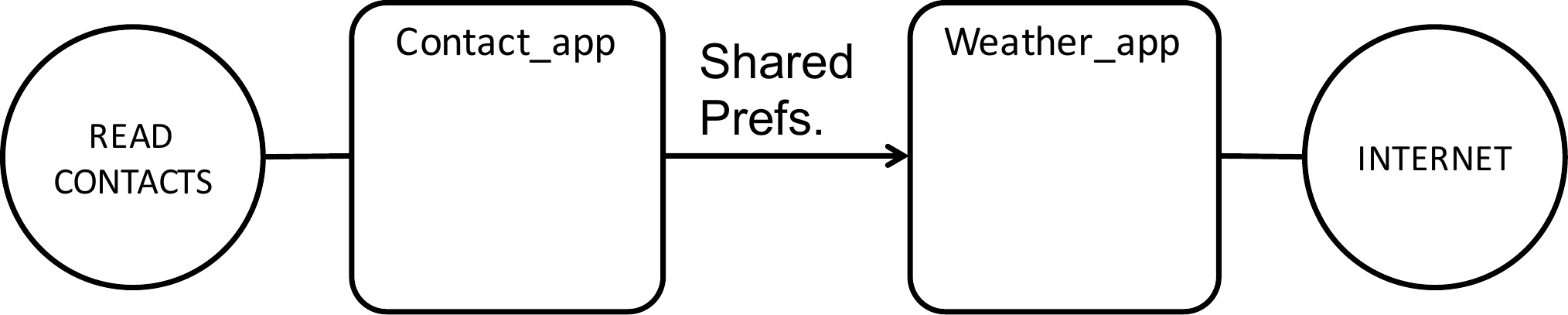}
\caption{An example of colluding apps}\label{fig:contact_stealer}
\end{figure}

To illustrate our definition we present an example.

\begin{example}[Stealing contact data]
The two apps graphically represented in Figure
\ref{fig:contact_stealer} perform information theft: the {\tt
  Contact\_app} reads the contacts database to pass the data to {\tt Weather\_app}, which sends the data to a remote server
controlled by the adversary. The information is sent through shared
preferences.

Using the collusion definition we can describe the actions performed
by both apps as:
$ Act_{\tt Contact\_app} = \{a_{\mathit read\_contacts} \}, $
$ Act_{\tt Weather\_app} = \{a_{\mathit send\_file}\}. $
with 
$\pms(a_{\mathit read\_contacts}) = \{ \mathit{Permission\_contacts} \}$
and
$\pms(a_{\mathit send\_file}) = \{ \mathit{Permission\_internet} \}.$
The information threat $t$ is given by 
$ T = \{ a_{\mathit read\_contacts}, a_{\mathit send\_file} \} $
and defining
$a_{\mathit read\_contacts} < a_{\mathit send\_file}.$
The inter-app communication is defined as 
$com_{\tt Contact\_app} = \{send_{\mathit shared\_prefs}\},$
$com_{\tt Weather\_app} = \{recv_{\mathit shared\_prefs}\} $
and
$send_{\mathit shared\_prefs} < recv_{\mathit shared\_prefs}. $

\end{example}

%% we suggest to define a metric, look for ``clusterpoints'' - unclear:
%% how to produce these?

%% Given a set of apps:

%% - Prolog program (London)
%% - Statistical analysis (Coventry)

%% result: 
%% a) Matrix with collusion suspicion (prolog), risk value (for pairs)
%% b) risk values for connected components of the prolog output
      
%% this gives indication for deeper analysis

%% deeper analysis: model checking in K

\section{Detecting collusion threat}\label{sec:prolog}

Our first approximation to detect app collusion utilises Logic Programming in Prolog. Its goal is to serve as a fast, computationally cheap filter that detects potential
colluding apps. It (1) uses Androguard~\cite{desnos2013androguard} to extract facts about
the communication channels and permissions of all single apps in a
given app set $S$, (2) which is then abstracted into an
over-approximation of actions and communication channels that could be used by a single app. (3) Finally the collusion rules are fired if the proper combinations of actions and communications are found in $S$.

%\subsection{Actions}

We utilise an action set $Act_{prolog}$ composed out of four different
high level actions: accessing sensitive information, use an API that
can directly cost money, control device services (e.g. camera, etc.),
and send information to other devices and the Internet. To find out
which of these actions an app could carry out, we extract its set of
permissions $pms_{prolog}$. Each permission is mapped to one or more
of the four high level actions. For example, an
app that declares the \lstinline|INTERNET| permission will be capable
of sending information outside the device:
\begin{equation*}
\mathit{uses}(App,P_{Internet}) \rightarrow \mathit{information}\_outside(App)
%\label{eq:internetpermission}
\end{equation*}

%\subsection{Communications}
The communication channels established by an app are characterised by
its API calls and the permissions declared in its manifest
file. We cover all communication actions 
($com_{prolog}$) that can be created as follows:
\begin{itemize}
\item \textit{Intents} are messages used to request actions from other
  application components (activities, services or broadcast
  receivers). 
  %% These can belong to the same or different
  %% apps. 
   Activities, services and broadcast receivers declare the
  intents they can handle by declaring a set of
  \lstinline|IntentFilters|.

\item \textit{External Storage} is a storage space shared between all
  the apps installed without restrictions. Apps accessing the external
  storage need to declare the \lstinline|READ_EXTERNAL_STORAGE|
  permission. To enable writing, apps must declare
  \lstinline|WRITE_EXTERNAL_STORAGE|.

\item \textit{Shared Preferences} are an OS feature to store key-value
  pairs of data. Although it is not intended for inter-app
  communication, apps can use key-value pairs to exchange information
  if proper permissions are defined (before Android 4.4).
\end{itemize}

We map apps to sending and receiving actions by inspecting their code
and manifest files. When using \lstinline|Intents| and
\lstinline|SharedPreferences| we are able to specify the communication
channel using the intent actions and preference file and package
respectively. If an application sends a
\lstinline|BroadcastIntent| with an action \lstinline|SEND_FILE| we
consider the following:
\begin{eqnarray*}
send\_broadcast(App,Intent_{send\_file}) \\
\rightarrow send(App,Intent_{send\_file})
\end{eqnarray*}
We consider that two apps communicate if one of them is able to $send$
and the other to $receive$ through the same channel. This allows to
detect communication paths composed by an arbitrary number of apps:
\begin{eqnarray*}
send(App_a,channel) \wedge receive(App_b,channel) \rightarrow   \\  
 communicate(App_a,App_b,channel)
\end{eqnarray*}
%\subsection{Threats}
Our threat set $\tau_{prolog}$ considers information theft, money theft and service misuse. 
As our definition states, each of the threats is characterised by a sequence of actions. For example, the information theft threat is codified as
the following Prolog rule:
$$
\begin{array}{lll}
            &        & \mathit{sensitive\_information}(App_a) \\
            & \wedge & \mathit{information\_outside}(App_b)  \\
            & \wedge & \mathit{communicate}(App_a,App_b,channel) \\ 
\rightarrow &        & \mathit{collusion}(App_a,App_b)
\end{array}
$$ 
Currently, we do not take into account the order of action
execution.

\section{Assessing the collusion possibility}\label{sec:coventry}

In this section, we apply machine learning to classify app sets into
colluding and non-colluding ones. To this end, we first define a
probabilistic model. Then we train the model, i.e., estimate the model
parameters on a training data set. As a third step we validate the
model using a validation data set. Additionally, in
Section~\ref{stat:testing}, we check the model with testing data.

%In many classification problems explicit rules do not exist or are rather difficult to define. Hence a classifier cannot be constructed from known rules and therefore one tries to infer a classifier from a (limited) set of training examples. On the other hand, inherent weakness associated with rule based approaches is its inability to capture the \emph{motivation} uncertainty behind an action. For example, SEND\_SMS can be used maliciously as well as legitimately needed by communication apps. Defining rules to capture this kind of uncertainty would be very expensive or sometimes not feasible. Though the context would be the key, we hypothesise that considering all possible actions in $S$ collectively (as a multivariate variable)  would give better results in terms of motivation than looking at individual actions. This section describes our approach to estimate potentially colluding app pairs using a probabilistic approach.

%\subsection{Probabilistic Model}
%{Computing $L_{\tau}$ and $L_{com}$}
Estimating the collusion possibility within a set $S$ of apps involves
to estimate two different likelihood components $L_{\tau}$ and
$L_{com}$. $L_{\tau}$ denotes the likelihood of carrying out a
threat. $L_{com}$ denotes the likelihood of performing some inter-app
communication. Hence, the likelihood of colluding within $S$ is given
by $L_{\tau} \times L_{com}$.

In order to estimate $L_{\tau}$, we employ a so-called Naive Bayesian
informative~\cite{PengNaive} model. We consider a multi-variate random variable $Y = (y_1, \dots, y_k)$. Here, $k$ is
the total number of permissions in Android OS, and $y_j \in
\{0,1\}$ are independent Bernoulli random variables. Variable $y_j$ is 1 if permission $j$ is found in $S$, 0 otherwise. $P(Y)$ stands for the probability of obtaining $S$ with
permissions as described by $Y$. Our probabilistic model is then given
by equation (\ref{Naive_Bayesian})
\begin{equation}
\label{Naive_Bayesian}
P(Y)=\prod_{j=1}^{k}\lambda_{j}^{y_{j}}(1-\lambda _{j})^{1-y_{j}}
\end{equation}
where $\lambda_j \in [0,1] $ is the Bernoulli parameter.

To compute $L_{\tau}$ we define $L_\tau=ln \{(P(Y))^{-1}\}$. This reflects on the likelihood of an app set to ``being malicious'' increasing as the number of permissions requested increase. Therefore, any monotonically decreasing function can be chosen~\cite{PengNaive}. For
comparisons we average out $L_\tau$ and scale down to the range [0,1].

To complete our modelling, we need to estimate values $\hat{\lambda
  _{j}}$ that replace $\lambda _{j}$ in the computation of $L_\tau.$
To this end -- to avoid over fitting $P(Y)$ -- we estimate $\lambda
_{j}$ using informative beta prior
distributions~\cite{Krishnamoorthy2015} and define the maximum
posterior estimation
\begin{equation}
\label{lamdahat}
\hat{\lambda _{j}}=\frac{\sum y_{j}+ \alpha_{j}}{N+ \alpha_{j}+\beta_{j}}
\end{equation}
where $N$ is the number of apps in the training set and
$\alpha_j, \beta_j$ are penalty effects. In this work we set
$\alpha_j =1.$ $\beta_j$ depend on the critical level
of permissions as given in~\cite{SarmaRisk,PengNaive}. $\beta_j$ can
take either the value $2N$ (most critical), $N$ (critical) or $1$
(non-critical).

We consider $L_{com}$ to be a binary function such that $L_{com} \in
\{1,0\}$ which takes the value 1 if there is inter app communication
within $S$, 0 otherwise.

%\subsection{Model Training}

To the best of our knowledge, there are no known colluding examples in
the wild. However, there are sets of apps available, where individual
apps have been classified by industry experts.  Thus, we have utilised
one such set provided by Intel Security, which consists of 9k+
malicious and 9k+ clean apps. As mentioned in Section 2, there is no
evidence suggesting differences between threats caused by single apps
and colluding apps. Thus, we can estimate $L_\tau$ based on this
set.
%%  It explains the fact that to what extend the set of permissions
%% in S can do a malicious activity if they were in a single app.
As for
$L_{com},$ there is no need to estimate any constants.

%Based on this app set, we estimate $\hat{\lambda _j}$ as follows:

%Let $F_\tau$ denote the set of all permissions required to describe a threat in $\tau$. Let $F_{com}$ denote the set of features required for  inter app communication. We will use $F_\tau$ in ordr to estimate $L_\tau$, while $F_{com}$ is used to estimate $L_{com}$.  

%So we assume that $F_\tau$ would be similar in both cases, and it allows us to use single apps in estimating $L_\tau$. 

%\subsection{Model Validation}

\begin{figure}
\centering
\includegraphics[scale=0.5]{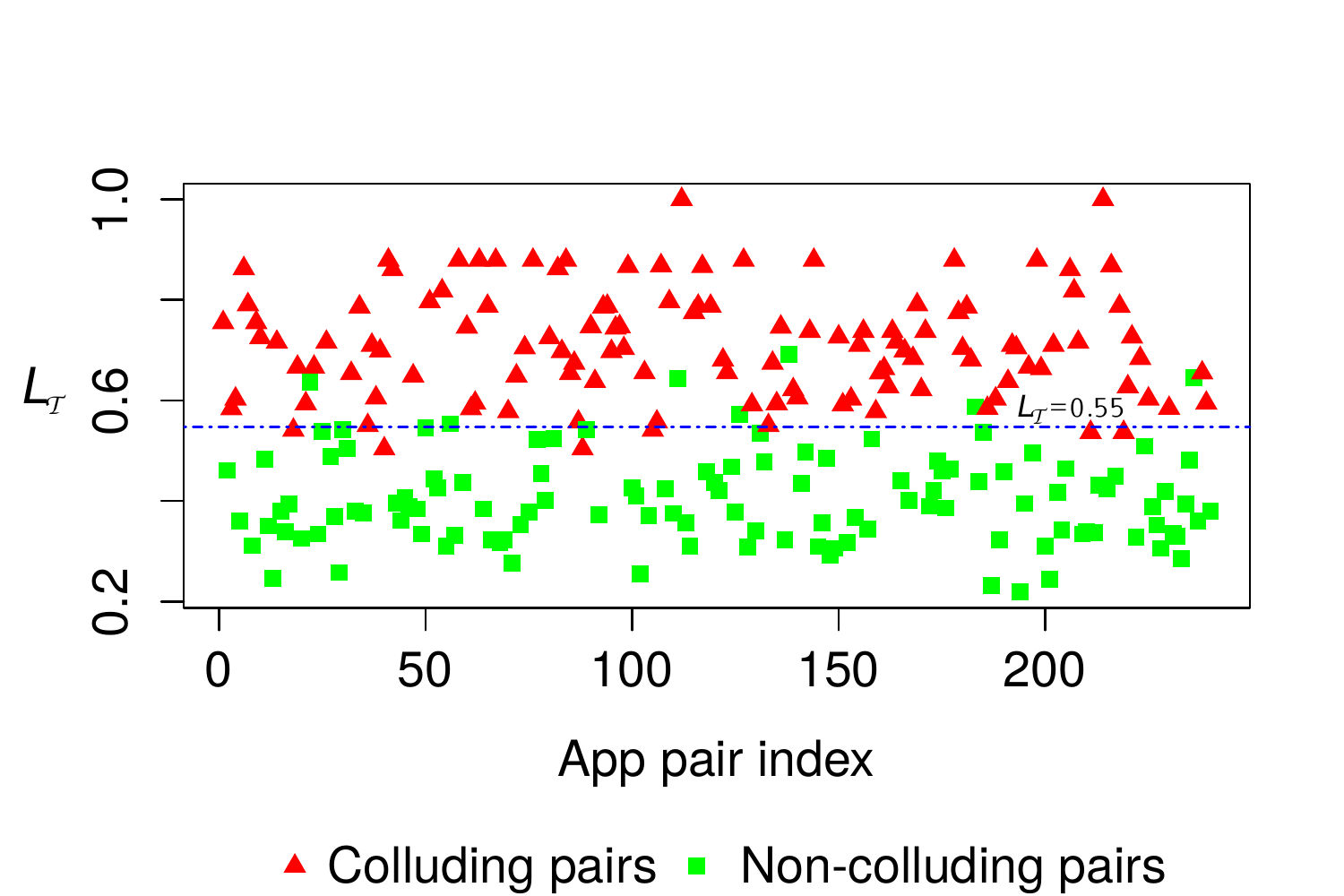}
\caption{Distribution of $L_{\tau}$ over evaluation sample. Blue
  dotted line: best possible linear discriminant line.}
\label{fig:evaluation}
\end{figure}

We validate our model on a larger set of apps (a blind sample) that
was produced for calibration of collusion detection mechanisms. This
validation set consists of 240 app pairs in which half are colluding
pairs, the other half is non-colluding.

For this analysis, we implemented an automated process using R and
Bash scripts, which also includes calls to a third party research
prototype~\cite{BurketdidFail} to find intent based
communications. Additionally, a simple permission based rule set was
defined to find communication using external storage. Overall, this
process consists of the following steps: (1) extracting permissions of
all single apps in a given app set S; (2) computing $L_{\tau}$ using
extracted permissions; (3) if $L_{\tau}$ is greater than a certain
threshold then estimating $L_{com}$ as mentioned above; and finally
(4) computing $L_{\tau} \times L_{com}.$

Figure~\ref{fig:evaluation} presents $L_{\tau}$ values for the
validation dataset in which a clear separation can be seen between two
classes with a lower (=0.50) and upper (=0.69) bounds for a
discriminant line.  Table~\ref{Table:confusionMatrixNaive} presents
the confusion matrix obtained by fitting the best possible linear
discriminant line at $L_{\tau}=0.55$ in Figure~\ref{fig:evaluation}.

\begin{table}[h]
\centering
{\footnotesize
\begin{tabular}{c|c|c|}
\cline{2-3}
{n=240}                                                                                    & {\begin{tabular}[c]{@{}c@{}}Actual\\ Colluding\end{tabular}} & {\begin{tabular}[c]{@{}c@{}}Actual\\ Non-Colluding\end{tabular}} \\ \hline
\multicolumn{1}{|c|}{{\begin{tabular}[c]{@{}c@{}}Predicted\\ Colluding\end{tabular}}}      & 114                                                                  & 7                                                                       \\ \hline
\multicolumn{1}{|c|}{{\begin{tabular}[c]{@{}c@{}}Predicted\\ Non-Colluding\end{tabular}}} & 6                                                                  & 113                                                                       \\ \hline
\end{tabular}
\caption{Confusion matrix for the evaluation sample.}
\label{Table:confusionMatrixNaive}
}
\end{table}

Performance measures precision(=0.94) and F-measure(=0.95) were
computed using the Table~\ref{Table:confusionMatrixNaive}. Precision
quantifies the notion of specificity while F-measure provides a way to
judge the overall performance of the model for each class. The higher
these values, the better the performance. Such values could be due to a bias of the validation sample towards the methodology. Further investigation on the bias of the evaluation dataset is left for future work.

\section{Model-Checking for collusion}\label{sec:swansea}

\begin{figure}
\centering
\includegraphics[height=2cm,width=8cm]{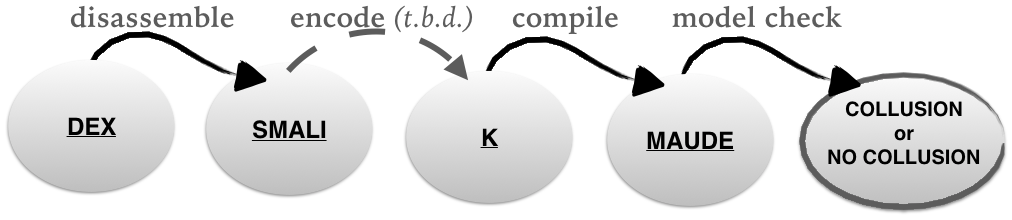}
\caption{Work-flow}\label{fig:mc_workflow}
\end{figure}

We demonstrate the effective attempt to detect collusion via model-checking. Figure
\ref{fig:mc_workflow} shows the basic work-flow employed: it starts
with a set of apps in the Dalvik Executable format DEX \cite{dex};
this code is disassembled and (currently manually) translated into a
semantically faithful representation in the framework K \cite{rosu15}
-- this step also triggers a data flow analysis of the app code;
compilation translates the K representation into a rewrite theory in
Maude \cite{maude}; using the Maude model-checker provides an answer
if the app set colludes or not: in case of collusion, the tool
provides a readable counter example trace (see Section
\ref{ssec:dataflowExp}). 

The rewriting logic semantics framework K allows the user to define
\emph{configurations}, \emph{computations} and \emph{rules} \cite{rosu15}. \emph{Configurations}
organise the program state in units called cells. In case of SMALI
assembly programs, on the method level program states essentially
consist of the current instruction, registers and parameters, and the
class it belongs to. The operational semantics of, say, the assembly
instruction
\begin{center}
{\tt const} {\sl Register, Int} 
\end{center}
for loading an integer constant {\sl Int} into a register {\sl
  Register} is captured by a \emph{rule}
\begin{center}
\includegraphics[height=1.7cm]{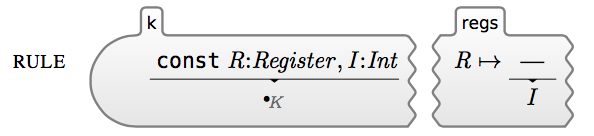}
\end{center}
which reads: provided the current instruction is to load an integer
{\sl I} into a register {\sl R}, then the cell ``{\sl regs}''
capturing the state of the registers is updated by binding the value
{\sl I} to register {\sl R}. The SMALI language also includes
procedure calls. These allow to access functionality provided by the
Android operating system, e.g., to read protected resources such as
the GPS location or to send/receive a broadcast intent. Thus, besides
encoding SMALI instructions in K, we also model the infrastructure
that Android provides to apps.

%{\bf question to Irina: this should run under ``computation''?}

The executions of the above semantics, instruction after instruction,
defines the \emph{computations}. However, the concrete semantics is
far too detailed for effective model-checking. Therefore, we implement
an abstraction in form of a data flow analysis. Here we represent each
method individually as a graph, where registers, types, and constants
are the nodes, and there is an edge between two nodes $n_1$ and $n_2$
iff $n_2$ is a parameter of the SMALI command that stores a value in
$n_1$. For example, the SMALI command {\tt const} {\sl r1, 42} would
lead to an edge from node ``42'' to node ``{\sl R1}''. Analysing such
graphs allows to detect which commands influence the values sent, say,
in a broadcast intent, or publishing via the internet; these commands
can be grouped into blocks and -- rather than computing with concrete
values -- their effect can be captured symbolically.

For the detection of collusion, our K semantics carries a ``trace''
cell that records which operations provided by the Android API -- see
Definition \ref{def:collusion} -- have been executed in a specific
run. Based on the ``trace'' cell we define collusion via a K rule:
provided that the GPS location has been accessed, this value has been
sent in a broadcast, this value has been received from a broadcast,
and finally this value has been published, in that case we detect
collusion ``information theft''.

Practical experiments with small apps demonstrate that this approach
is feasible. Using the Maude model-checker, the state space of the
(abstraction) of two apps is small, with only 8 states for the given example and the check takes less than a second. %When modifying the apps in such a way that the information flow is broken, by, say using a different name for the broadcast and thus disabling communication between the apps, model-checking shows that no collusion is happening.

We use symbolic analysis over the byte code of the set S in order to
obtain an in-depth inspection of the communication patterns. In the byte code of each app in S we detect the flow of communication with another app and a safe
over-approximation of the data being communicated. %% As such, we
%% detect if any of the private data of the device is read by an
%% application having the adequate permissions but is also
%% distributed to other application(s) not having the required
%% permissions.
We use static analysis over the byte code of apps to extract communication-flow between apps and data-flow inside each app. The data-flow information is filtered based on the private data being read and communicated.

It is future work, to complete the encoding to cover the full DEX
language and to experiment with larger apps. We further intend to prove the abstraction to be sound, to show that if model-checking the abstracted code detects collusion then there is collusion in the original code, and to have a more general definition of collusion that covers different variants.

\section{First experimental results}
\label{sec:results}

In this section we compile experimental results obtained by applying the three methods discussed above to an artificial set.

\subsection{An artificial colluding app set}\label{ssec:artificial}

In order to validate our different approaches against a known ground
truth, we have created fourteen apps
%
%% We
%%have decided to use apps developed by us, instead using
%% already available apps, for two reasons. First, to the best of the
%% authors knowledge, no colluding apps have been identified in the
%% wild. Second, even when there are yet apps to be identified as being
%% colluding, we are not 100\% certain that an app downloaded from the
%% Google Play Store is not colluding, even though the app may be coming
%% from a trusted developer. Colluding apps have been developed to
%
that cover all threats and use all communication channels described
earlier in this paper. They are organised in four colluding sets:
The \textbf{Document Extractor} set consists of one app (\textit{id
  1}) that looks for sensitive documents on the external storage; the
other app (\textit{id 2}) sends the information received (via
\lstinline|SharedPreferences|) to a remote server.  The
\textbf{Botnet} set consists of four apps. One app (\textit{id 3})
acts as a relay that receives orders from the command and control
center. The other colluding apps execute commands (delivered via
\lstinline|BroadcastIntents|) depending on their permissions: sending
SMS messages (\textit{id 4}), stealing the user's contacts (\textit{id
  5}) and starting and stopping tasks (\textit{id 6}).  The
\textbf{Contact Extractor} set consists of three apps. The first
(\textit{id 7}) reads contacts from the address book, the second
(\textit{id 8}) forwards them via the external storage to the third
one (\textit{id 9}), which sends them to the Internet. The first and
second app communicate via \lstinline|BroadcastIntents|.  The
\textbf{Location Stealing} set consists of one app (\textit{id 12})
that reads the user location and shares it with the second app
(\textit{id 13}), which sends the information to the Internet.

The three non-colluding apps are a document viewer (\textit{id 10}),
an information sharing app (\textit{id 11}) and a location viewer
(\textit{id 14}). The first app is able to display different file
types in the device screen and use other apps (via
\lstinline|BroadcastIntents|) to share their uniform resource
identifier (URI). The second app receives text fragments from other
apps and sends them to a remote server. The third app receives a
location from another app (with the same intent used by apps 12 and
13) and shows it to the user on the screen.

\subsection{Detecting collusion threat with Prolog}

\begin{table}[t]
\centering
\setlength\tabcolsep{3pt}
\label{tab:col_matrix}
{\footnotesize
\begin{tabular}{| c | c | c |c | c|c | c | c | c | c | c | c | c | c| c |}
 \hline
\textbf{id} & \textbf{1} & \textbf{2} & \textbf{3} & \textbf{4} & \textbf{5} & \textbf{6} & \textbf{7} & \textbf{8} & \textbf{9} & \textbf{10} & \textbf{11} & \textbf{12} & \textbf{13} & \textbf{14} \\
 \hline
\textbf{1} &  & \textcolor{truepositive}{$\clubsuit$} & & & & & & & & \textcolor{falsepositive}{$\clubsuit$} & \textcolor{falsepositive}{$\clubsuit$} & & &  \\
  \hline
\textbf{2} & &  & & & & & & & & & & & &  \\
  \hline
\textbf{3} & & &  & \textcolor{truepositive}{\textdollar}\textcolor{falsepositive}{$\clubsuit$} & \textcolor{truepositive}{$\spadesuit$} & \textcolor{truepositive}{$\spadesuit$} & & & & & & & &  \\
  \hline
\textbf{4} & & & &  & & & & & & & & & &  \\
  \hline
\textbf{5}  & & & \textcolor{truepositive}{$\clubsuit$} &  &  & & & & & \textcolor{falsepositive}{$\clubsuit$} & \textcolor{falsepositive}{$\clubsuit$} & & &   \\
  \hline
\textbf{6}  & & & \textcolor{truepositive}{$\clubsuit$} & \textcolor{falsepositive}{$\clubsuit$} &  &  & & & & & & & &  \\
  \hline
\textbf{7} & & \textcolor{falsepositive}{$\clubsuit$} & & & & &  & \textcolor{truepositive}{$\clubsuit$} & \textcolor{truepositive}{$\clubsuit$} & \textcolor{falsepositive}{$\clubsuit$} & \textcolor{falsepositive}{$\clubsuit$} & & &  \\
  \hline
\textbf{8} & & & & & & & &  & \textcolor{truepositive}{$\clubsuit$} & & & & &   \\
  \hline
\textbf{9} & & & & & & & & & & & & & &  \\
\hline
\textbf{10} & & & & & & & & & & & \textcolor{falsepositive}{$\clubsuit$} & & &  \\
  \hline
\textbf{11} & & & & & & & & & & & & & &  \\
  \hline
\textbf{12} & & & & & & & & & & & & & \textcolor{truepositive}{$\clubsuit$} & \textcolor{falsepositive}{$\clubsuit$}  \\
  \hline
\textbf{13} & & & & & & & & & & & & & &  \\
  \hline
\textbf{14} & & & & & & & & & & & & & &  \\
  \hline
\end{tabular}
}
\caption{Collusion Matrix of the Prolog program. \textcolor{truepositive}{$\clubsuit$} = Information theft. \textcolor{truepositive}{\textdollar} = Money theft. \textcolor{truepositive}{$\spadesuit$} = Service misuse. \textcolor{falsepositive}{$\clubsuit$}, \textcolor{falsepositive}{\textdollar}, \textcolor{falsepositive}{$\spadesuit$} = False positives.}
\end{table}

Table \ref{tab:col_matrix} shows the results obtained with our ruled
based approached from Section \ref{sec:prolog} on
the apps from Section \ref{ssec:artificial}. The
entry ``dark red club'' in row 1, column 2 means: the program detects that app \textit{id 1} sends information to app
\textit{id 2}, and these two apps collude on an ``information
theft''. As we take communication direction into consideration, the
resulting matrix is non-symmetric, e.g., there is no entry in row 2,
column 1. The entry ``light red club'' in row 1, column 10 means: the program flags collusion of type ``information theft'' though the set \{\textit{id 1}, \textit{id 10}\} is clean.%% This provides further information about the
%% collusion attack. For instance, one can see the information leak in
%% Information theft attacks. Additionally, the way we defined the
%% communication rules makes it possible to identify transitive collusion %% attacks (i.e. app 7 colluding with app 9 through app 8). 
The approach identifies all colluding app sets. It also flags
eight false positives due to over-approximation.  Note, that there are
no false negatives due to the nature of our test set: it utilises only
those communication methods that our Prolog approach is able to
identify.%, i.e., our test set is biased.

%% . i.e., 
%% collusion potential but are not really colluding (i.e. false
%% negatives). This is mainly because two reasons. First, we do not
%% consider in our initial classification some of the communication
%% channels that are already widely use by apps in Android. For example,
%% the \lstinline|Intent| with action \lstinline|VIEW| or
%% \lstinline|SEND| are very common in Android applications. It is
%% unlikely that apps would use them for collusion as other apps could
%% have registered to receive the same information.Second, in this
%% approach, we identify apps that are communicating by sharing access to
%% sensitive resources, but we do not look at how that access is
%% shared. It must be noted, that the main aim of this approach is to
%% reduce the amount of app combinations that are being passed through
%% the data-flow analysis. 

%% The results shown in Table
%% \ref{tab:col_matrix} don't show any false negatives. This is because
%% the app set used for evaluation considers the same communication
%% methods as the ones the Prolog approach is able to
%% identify. Therefore, the results given by this approach need to be
%% considered in relation with the communication channels that are being
%% analysed. In order to reason about other communication channels, it
%% would be necessary to add the corresponding detectors (in addition to
%% the \lstinline|Intent|, \lstinline|SharedPreferences| and external
%% storage detectors).

\subsection{Computing collusion possibility}
\label{stat:testing}

\begin{table*}
\centering
\label{tab:col_matrix_satat}
\centering
{\tiny
\resizebox{2\columnwidth}{!}{%
\begin{tabular}{|l|l|l|l|l|l|l|l|l|l|l|l|l|l|l|}
\hline
\textbf{ID} & \textbf{1} & \textbf{2}                   & \textbf{3}                   & \textbf{4}                   & \textbf{5}                   & \textbf{6}                   & \textbf{7}                   & \textbf{8}                   & \textbf{9}                   & \textbf{10}                  & \textbf{11}                  & \textbf{12}                  & \textbf{13}                  & \textbf{14}                  \\ \hline
\textbf{1}  &            & \cellcolor[HTML]{9AFF99}0.51 & \cellcolor[HTML]{009901}0.61 & \cellcolor[HTML]{009901}0.97 & \cellcolor[HTML]{009901}1    & \cellcolor[HTML]{009901}0.8  & \cellcolor[HTML]{009901}1    & \cellcolor[HTML]{FFCCC9}0.81 & \cellcolor[HTML]{009901}0.77 & \cellcolor[HTML]{009901}0.77 & \cellcolor[HTML]{009901}0.77 & \cellcolor[HTML]{009901}0.44 & \cellcolor[HTML]{009901}0.44 & \cellcolor[HTML]{009901}0.95 \\ \hline
\textbf{2}  &            &                              & \cellcolor[HTML]{009901}0.48 & \cellcolor[HTML]{009901}0.62 & \cellcolor[HTML]{009901}0.55 & \cellcolor[HTML]{009901}0.49 & \cellcolor[HTML]{009901}0.55 & \cellcolor[HTML]{FFCCC9}0.58 & \cellcolor[HTML]{FFCCC9}0.51 & \cellcolor[HTML]{FFCCC9}0.51 & \cellcolor[HTML]{009901}0.58 & \cellcolor[HTML]{009901}0.31 & \cellcolor[HTML]{009901}0.31 & \cellcolor[HTML]{009901}0.49 \\ \hline
\textbf{3}  &            &                              &                              & \cellcolor[HTML]{FE0000}0.69 & \cellcolor[HTML]{FE0000}0.64 & \cellcolor[HTML]{FE0000}0.56 & \cellcolor[HTML]{009901}0.64 & \cellcolor[HTML]{009901}0.48 & \cellcolor[HTML]{009901}0.61 & \cellcolor[HTML]{009901}0.61 & \cellcolor[HTML]{009901}0.72 & \cellcolor[HTML]{009901}0.41 & \cellcolor[HTML]{009901}0.41 & \cellcolor[HTML]{009901}0.58 \\ \hline
\textbf{4}  &            &                              &                              &                              & \cellcolor[HTML]{009901}1    & \cellcolor[HTML]{009901}0.84 & \cellcolor[HTML]{009901}1    & \cellcolor[HTML]{009901}0.85 & \cellcolor[HTML]{009901}0.71 & \cellcolor[HTML]{009901}0.71 & \cellcolor[HTML]{009901}0.82 & \cellcolor[HTML]{009901}0.56 & \cellcolor[HTML]{009901}0.56 & \cellcolor[HTML]{009901}0.95 \\ \hline
\textbf{5}  &            &                              &                              &                              &                              & \cellcolor[HTML]{009901}0.84 & \cellcolor[HTML]{009901}1    & \cellcolor[HTML]{009901}0.86 & \cellcolor[HTML]{009901}0.67 & \cellcolor[HTML]{009901}0.67 & \cellcolor[HTML]{009901}0.82 & \cellcolor[HTML]{009901}0.47 & \cellcolor[HTML]{009901}0.47 & \cellcolor[HTML]{009901}1    \\ \hline
\textbf{6}  &            &                              &                              &                              &                              &                              & \cellcolor[HTML]{009901}0.84 & \cellcolor[HTML]{009901}0.68 & \cellcolor[HTML]{009901}0.58 & \cellcolor[HTML]{009901}0.58 & \cellcolor[HTML]{009901}0.65 & \cellcolor[HTML]{009901}0.43 & \cellcolor[HTML]{009901}0.43 & \cellcolor[HTML]{009901}0.78 \\ \hline
\textbf{7}  &            &                              &                              &                              &                              &                              &                              & \cellcolor[HTML]{FE0000}0.86 & \cellcolor[HTML]{9AFF99}0.67 & \cellcolor[HTML]{009901}0.67 & \cellcolor[HTML]{009901}0.82 & \cellcolor[HTML]{009901}0.47 & \cellcolor[HTML]{009901}0.47 & \cellcolor[HTML]{009901}1    \\ \hline
\textbf{8}  &            &                              &                              &                              &                              &                              &                              &                              & \cellcolor[HTML]{FE0000}0.51 & \cellcolor[HTML]{FFCCC9}0.51 & \cellcolor[HTML]{009901}0.58 & \cellcolor[HTML]{009901}0.31 & \cellcolor[HTML]{009901}0.31 & \cellcolor[HTML]{009901}0.77 \\ \hline
\textbf{9}  &            &                              &                              &                              &                              &                              &                              &                              &                              & \cellcolor[HTML]{009901}0.77 & \cellcolor[HTML]{009901}0.77 & \cellcolor[HTML]{009901}0.44 & \cellcolor[HTML]{009901}0.44 & \cellcolor[HTML]{009901}0.61 \\ \hline
\textbf{10} &            &                              &                              &                              &                              &                              &                              &                              &                              &                              & \cellcolor[HTML]{009901}0.77 & \cellcolor[HTML]{009901}0.44 & \cellcolor[HTML]{009901}0.44 & \cellcolor[HTML]{009901}0.61 \\ \hline
\textbf{11} &            &                              &                              &                              &                              &                              &                              &                              &                              &                              &                              & \cellcolor[HTML]{009901}0.47 & \cellcolor[HTML]{009901}0.47 & \cellcolor[HTML]{009901}0.73 \\ \hline
\textbf{12} &            &                              &                              &                              &                              &                              &                              &                              &                              &                              &                              &                              & \cellcolor[HTML]{9AFF99}0.47 & \cellcolor[HTML]{009901}0.41 \\ \hline
\textbf{13} &            &                              &                              &                              &                              &                              &                              &                              &                              &                              &                              &                              &                              & \cellcolor[HTML]{009901}0.41 \\ \hline
\textbf{14} &            &                              &                              &                              &                              &                              &                              &                              &                              &                              &                              &                              &                              &                              \\ \hline
\end{tabular}%
}
}
\caption{Matrix for collusion possibility.}
\end{table*}

Table \ref{tab:col_matrix_satat} shows results from our
alternative approach from Section \ref{sec:coventry} on the same set of apps. %the app set described in Section \ref{ssec:artificial} above.  
Each cell denotes the $L_{\tau}$ value for the corresponding pair. To
minimise false negatives, we use the lower bound (=0.50) gained from
the validation data set for the discriminant line as threshold for
$L_{\tau}$. We report possible collusion if $L_{\tau}\geq 0.5$ and
$L_{com}=1$, otherwise we report non-collusion. This yields symmetric
data -- for readability we leave the lower half of the matrix  empty. Dark red shows true
positives, light red shows false positives, dark green shows true
negatives, and light green shows false negatives.

This approach finds two of the four colluding app sets and flags five false positives.  It relies on a 3rd party tool to detect inter-app communication which ignores communication using SharedPreferences, thus the app set
\{\textit{id 1}, \textit{id 2}\} is not detected. As we restrict
ourselves to pairwise analysis only, the app set \{\textit{id 7},
\textit{id 9}\} can't be detected, as it communicates via \textit{id
  8}. Finally, app set \{\textit{id 12}, \textit{id 13}\} was not
reported since its $L_{\tau}$ value is less than the chosen
threshold. Choosing a lower threshold could avoid this false negative,
but at the cost of a lower class accuracy and performance.

%% There is a possibility to happen the same collusion scenario described
%% for app pair (1,2) in pairs reported as false positives (except
%% (1,8)).  Because they have the same (or more) permissions as in pair
%% (1,2) and can communicate between two apps using external storage.  A
%% deep analysis is needed to decide this. We need to check communication
%% between methods in each app (intra app) in a pair to verify the
%% continuity of the communication channel. Knowing set of permissions in
%% S and finding an inter app communication is not enough for this
%% purpose and a detail static code/dynamic analysis is needed. This
%% would be the next step of our tool.

A precise estimation of $L_{com}$ would be useful in order to reduce
the number of false positives in our analysis. However, communication
is only a necessary, not a sufficient condition for collusion. A recent
study~\cite{ElishIAC} shows that 84.4\% of non-colluding apps in the
market place can communicate with other apps either using explicit
(11.3\%) or implicit (73.1\%) intent calls. Therefore the threat
element (i.e. $L_{\tau}$) is far more informative in collusion
estimation than the communication element ($L_{com}$) in our model.
%% The test sample is a blind sample and we have not properly
%% investigated it for the biasednes or realisticity.

Both approaches to detect the potential of collusion are constrained 
in terms of the type of inter-app communication channels they account 
for due to reasons explained previously. This makes it difficult to 
provide for a straightforward comparison. The rule-based approach is 
not limited to pairs of set for collusion (which may involve more than 
two apps). It also allows us assess for the direction of the colluding 
behaviour (for cases of information flow). Defining rules requires 
expert knowledge and for some cases explicit rules may not exist; this 
is overcome by providing for rules to over-approximate for potentially 
colluding behaviour.

A statistical approach, on the other hand, has the advantage that it 
can reflect on varying degrees of possible collusion for a given set 
of apps. In this paper we have largely focused on static attributes, 
such as permissions, a number of other similar static attributes (such 
as developer ID, download metrics, and so on) may also be placed on a 
scale of suspicion for collusion; albeit the current implementation of 
the statistical approach in this paper is limited due to tool 
availability.

\subsection{Software model-checking}\label{ssec:dataflowExp}

We inspect two sets of applications, one colluding \{\textit{id 12},
\textit{id 13}\} and another non-clouding pair \{\textit{id 12},
\textit{id 14}\} with software model checking as described in Section
\ref{sec:swansea}.
%
%In these two pairs we have a sender (with permissions to access the GPS location) and a receiver (which does not have permissions to access the GPS location). 
%
The sender \textit{id 12} and receiver (\textit{id 13} or \textit{id
  14}) communicate via broadcast messages containing the details of
the GPS location. In the colluding case the broadcast is successful,
i.e., the sender reads the GPS location then broadcasts it, and the
receiver gets the broadcast then publishes the GPS location on the
internet. In the non-colluding case, the sender still broadcasts the
private data which reaches the receiver but this data is never
published. Instead, the receiver publishes something else. The
data-flow analysis shows, for the non-colluding case, that the private
data is received but is never published. This aspect is not obvious
for our Prolog filter. Similarly, the data-flow analysis detects
collusion in the first case and reports its path witnesses, such as:

\noindent
\texttt{<trace>} 
$\ $ \texttt{call(readSecret, p1)} \\
$\ $ \texttt{-> r1 := callRet(readSecret)}\\
$\ $ \texttt{-> call(getBroadcast,r1,r1,"locsteal",p1)}\\
$\ $ \texttt{-> call(sendBroadcast,"locsteal",r1)}\\
$\ $ \texttt{-> r2 := callRet(getBroadcast)}\\
$\ $ \texttt{-> call(publish,r2)}\\
\texttt{</trace>}

\section{Related work}\label{sec:related}

App collusion can be traced back to \emph{confused deputy} attacks \cite{hardy1988confused}. They happen in form of \emph{permission re-delegation attacks} \cite{felt2011permission,davi2010privilege,wu2015effective} when a permission is carelessly exposed through a public component. %This happens when a third party abuses an application that is exposing a restricted resource through a public interface. %anotherA condition for this attack to happen is that there must exist an application that provides a public interface to access some restricted resources. If this happens, another application may use that interface to abuse the restricted resources. The application providing access to the protected/restricted resource is called a confused deputy. 
Soundcomber \cite{schlegel2011soundcomber} %% . This proof of concept is composed by two apps. The first app uses its access to the microphone to listen for bank credentials when the user calls telephone banking services. The second app transmits the stolen information to a remote server.
is an example where extracted information is transmitted using both overt and covert channels.

%The first papers to generalize the idea of permission re-delegation to create a collusion attack were \cite{bugiel2011xmandroid,marforio2011application}. 
Marforio et al.\ \cite{marforio2011application} define colluding applications as
those applications that cooperate in a violation of some security
property of the system. Another definition is given by \cite{mallcritical} where multiple apps can come together to perform a certain task, which is out of their
permission capabilities. The malicious component of collusion is also acknowledged by Bagheri, Sadeghi et al.~\cite{sadeghi2015analysis}. A more detailed definition is given by Elish \cite{ElishIAC}, which defines it as the collaboration between malicious apps, likely written by the same adversary, to obtain a set of permissions to perform attacks.

ComDroid \cite{chin2011analyzing} is a static analysis tool that looks
for confused deputies through \emph{Intents}. XManDroid
\cite{bugiel2011xmandroid} and TrustDroid \cite{bugiel2011practical}
extend the Android OS. Both allow for fine-grained policies that control inter-app information exchange; none of them address covert channels. In \cite{suarezcompartmentation} authors analyze,
using different risk metrics, several compartmentalisation strategies
to minimise the risk of app collusion, showing two or
three app compartments drastically reduce the risk of collusion for a
set of 20 to 50 apps.

\section{Summary}\label{sec:summary}

We have presented a new, concise definition of collusion in the
context of Android OS. Colluding apps may carry out information theft, money theft, or service misuse; to this end,
malware is distributed over multiple apps. As demonstrated by Soundcomber (and our
own app set), collusion is a possibility -- but it is yet unclear
whether it exists in the wild. Together with our industrial partner
Intel Security, we have developed a number of approaches towards effectively detecting collusion. Early experimental results
on small app sets (of a size nearly the scale of the number of apps
one has on a phone) look promising: a combination of the rule based
approach and the machine learning approach could serve as a filter, after which we employ model checking as a decision
procedure.

\section{Future work}\label{sec:futurework}

A frontal attack on detecting collusions to analyse analysing pairs, triplets and larger sets is not practical given the search space.
%due to the sheer amount of combinations of publicly available apps. Thus, 
%% we consider the step of pre-filtering apps essential for a
%% collusion detection system if it were to be used in practice.  %% Even if
%% we could find all collusions in all existing apps, new ones appear
%% every day and they could create new collusions with previously
%% analyzed apps. Continuously re-analysing the growing space of all
%% Android apps is infeasible so
An effective collusion-discovery tool must include an
 effective set of methods to isolate potential sets which require further examination.
% are deemed ``interesting''.

We have to emphasise that such tools are essential for many
collusion-detection methods. Besides our model checking
approach, one could, for example, think of the approach to merge pair
of apps into a single app to inspect aggregated data flows \cite{apkcombiner2015}. Merging is slow and therefore app-combining approach is predicated on effective filtering of app pairs of interest.

%% In this paper we researched a filter based on Android app permission declarations.
%% A filter based on Android app permissions is the simplest one. Additionally, 
%% permissions are very easy and cheap to extract from APKs -- no de-compilation, 
%% reverse engineering, complex code or data flow analysis is required.

Alternatively (or additionally), to the two filters described in our
paper, imprecise heuristic methods to find ``interesting'' app sets
may include: statistical code analysis of apps (e.g. to locate APIs
potentially responsible to communication, accessing sensitive
information, etc.); and taking into account apps' publication time and
distribution channel (app market, direct installation, etc.).

Attackers are more likely
to release colluding apps in a relatively short time frame and that
they are likely to engineer the distribution in such a way that
sufficient number of users would install the whole set (likely from
the same app market). To discover such scenarios one can employ:
analysis of security telemetry focused on users’ devices to examine installation/removal of apps, list of processes simultaneously executing, device-specific APK download/installation logs from app markets (like Google Play\texttrademark) and meta-data about APKs in app markets (upload time by developers, developer ID, source IP, etc.).
Such data would allow constructing a full view of existing app sets on
user devices. Only naturally occurring sets (either installed on same device or actually executing simultaneously) may be analysed for collusion which should drastically reduce the number of sets that require deeper analysis.

%Naturally, finding ``interesting'' app sets is not enough: in the end, some analysis is required to figure out if a given set of apps colludes. Manual analysis is costly, merging apps into a single one often fails, however software model checking of suitable abstractions of an app set might be a way forward.

Our goal is to build a fully automated and effective collusion
detection system, and tool performance will be central to address scale.  It is not clear yet where the bottleneck will be when we apply our approach to real-life apps. Further work will focus on identifying these bottlenecks to optimise the slowest elements of our tool-chain. Detecting covert
channels would be a challenge as modelling such will not be trivial. 
%However, this will be a natural step in the future evolution of the project.

%From an even more long-term perspective, collusions are a part of a general problem of effective isolation of software. This problem exists in all environments which implement sandboxing of software --from other mobile operating systems (like iOS and Tizen) to virtual machines in server farms (like Amazon EC2, Microsoft Azure and similar). We can see how covert communications between sandboxes may be used to breach security and create data leaks. The tendency to have more and better isolation is, of course, a positive one but we should fully expect the attackers to employ collusion methods more often to circumvent security. We endeavor to see if our methods developed for Android would be applicable to a wider range of operating environments.

\section*{Acknowledgement}
% Should mention grant numbers? 
This work has been funded by EPSRC and we are excited to work on this challenging piece of research\footnote{All code
  related to the project can be found in our GitHub page:
  \url{https://www.github.com/acidrepo}}.

\bibliographystyle{abbrv} 
\bibliography{main}
\end{document}